\documentclass{article}
\usepackage{ijcai18}

\usepackage{xcolor}
\usepackage{soul}
\usepackage[utf8]{inputenc}
\usepackage[small]{caption}
\usepackage{times}  %Required
\usepackage{helvet}  %Required
\usepackage{courier}  %Required
\usepackage{url}  %Required
\usepackage{graphicx}  %Required
\usepackage[utf8]{inputenc}
\usepackage[english]{babel}
\usepackage{amsmath}
\usepackage{amssymb}
\usepackage{amsthm}
\usepackage{bm}
\usepackage{multirow}		
\usepackage{calligra}	
\usepackage{epstopdf}	
\usepackage{subcaption}% http://ctan.org/pkg/subcaption
\title{LCMR: Local and Centralized Memories for Collaborative Filtering with Unstructured Text}

\author{
Guangneng Hu,
Yu Zhang,
Qiang Yang
\\
HKUST \\
njuhgn@gmail.com,
yuzhangcse@ust.hk,
qyang@cse.ust.hk
}
\begin{document}

\maketitle

\begin{abstract}
  Collaborative filtering (CF) is the key technique for recommender systems. Pure CF approaches exploit the user-item interaction data (e.g., clicks, likes, and views) only and suffer from the sparsity issue. Items are usually associated with content information such as unstructured text (e.g., abstracts of articles and reviews of products). CF can be extended to leverage text. In this paper, we develop a unified neural framework to exploit interaction data and content information seamlessly. The proposed framework, called LCMR, is based on memory networks and consists of local and centralized memories for exploiting content information and interaction data, respectively. By modeling content information as local memories, LCMR attentively learns what to exploit with the guidance of user-item interaction. On real-world datasets, LCMR shows better performance by comparing with various baselines in terms of the hit ratio and NDCG metrics. We further conduct analyses to understand how local and centralized memories work for the proposed framework.
\end{abstract}
	
\section{Introduction}

Recommender systems are widely used in various domains and e-commerce platforms,
such as to help consumers buy products at Amazon, watch videos on Youtube, and read articles on Google News. Collaborative filtering (CF) is among the most effective approaches based on the simple intuition that if users rated items similarly in the past then they are likely to rate items similarly in the future~\cite{sarwar2001item}. Matrix factorization (MF) techniques which can learn the latent factors for users and items are its main cornerstone~\cite{mnih2008probabilistic,koren2009matrix}. Recently, neural networks like multilayer perceptron (MLP) are used to learn the interaction function from data~\cite{dziugaite2015neural,he2017neural}. MF and neural CF suffer from the data sparsity and cold-start issues.

Items are usually associated with content information such as unstructured text, like the news article and product reviews. These additional sources of information can alleviate the sparsity issue and are essential for recommendation beyond user-item interaction data. For application domains like recommending research papers and news articles, the unstructured text associated with the item is its text content~\cite{wang2011collaborative,bansal2016ask}. Other domains like recommending products, the unstructured text associated with the item is its user reviews which justify the rating behavior~\cite{mcauley2013hidden,zheng2017joint}. Topic modelling and neural networks have been proposed to exploit the item content leading to performance improvement.

These two research threads, i.e., pure CF approaches which exploit user-item interaction data and extended CF methods which integrate item content, are different perspectives of interaction and content information. On the one hand, a recent pure CF approach called latent relational metric learning model~\cite{Tay2018latent} showed that user-item specific latent relations can be generated by a centralized memory module. The centralized (or global) memories are parameterized by a memory matrix shared by all user-item interaction data. This model is a pure CF approach. On the other hand, memory networks are widely used in question answering and reading comprehension~\cite{weston2016memory,sukhbaatar2015end,miller2016key}. The memories can be naturally used to model additional sources like item content. These memories are local (or dynamic) since they are specific to the input query. Based on local and centralized memories, there is a possibility of unifying these two research threads to exploit interaction data and content information seamlessly.

In this paper, we propose a novel neural framework to exploit interaction data and content information seamlessly from the centralized and local perspectives. The proposed framework, called LCMR, consists of local and centralized memories modules for exploiting content information and interaction data, respectively. By modeling content information as local memories, LCMR attentively learns what to exploit with the guidance of user-item interaction. The local and centralized memories are jointly trained under the end-to-end neural framework. Moreover, LCMR is a unified framework embracing pure CF and extended CF approaches.

\section{The LCMR Framework}

We first introduce notations used throughout the paper. For collaborative filtering with implicit feedback~\cite{hu2008collaborative,pan2008one}, there is a binary matrix $\bm{R} \in \mathbb{R}^{m \times n}$ to describe user-item interactions where each entry $r_{ui} \in \{0,1\}$ is 1 (called observed entries) if user $u$ has an interaction with item $i$ and 0 (unobserved) otherwise:
\[
r_{ui} =\left\{
        \begin{array}{ll}
          1, \quad \textrm{if interaction (user,item) exists}; \\
          0, \quad\textrm{otherwise}.
        \end{array}
      \right.
\]
Denote the set of $m$-sized users by $\mathcal{U}$ and $n$ items by $\mathcal{I}$. Usually the interaction matrix is very sparse since a user $u \in \mathcal{U}$ only consumed a very small subset $\mathcal{I}_u^+$ of all items. Similarly for the task of item recommendation, each user is only interested in identifying top-$N$ items. The items are ranked by their predicted scores:
\begin{equation}
\hat r_{ui} = f(u,i | \Theta),
\end{equation}
where $f$ is the interaction function and $\Theta$ model parameters.

For MF-based CF approaches, the interaction function $f$ is fixed and computed by a dot product between the user and item vectors~\cite{mnih2008probabilistic,koren2009matrix}. For neural CF, neural networks are used to parameterize function $f$ and learn it from interaction data~\cite{dziugaite2015neural,he2017neural}:
\begin{equation}
f(\bm{x}_{ui} | \bm{P}, \bm{Q}, \theta_f) = \phi_o( \phi_L(...(\phi_1(\bm{x}_{ui}))...)),
\end{equation}
where the input $\bm{x}_{ui} = [\bm{P}^T\bm{x}_u, \bm{Q}^T\bm{x}_i] \in \mathbb{R}^d$ is vertically concatenated from that of user and item embeddings, which are projections of their one-hot encodings $\bm{x}_u \in \{0,1\}^m$ and $\bm{x}_i \in \{0,1\}^n$ by embedding matrices $\bm{P} \in \mathbb{R}^{m \times d_1}$ and $\bm{Q} \in \mathbb{R}^{n \times d_2}$, respectively ($d = d_1 + d_2$). The output and hidden layers are computed by $\phi_o$ and $\{\phi_l\}$ in a multilayer neural network.

Items are associated with unstructured text like abstracts of articles and reviews of products. For item $i$ (such as an article), denote the $n_i$ words in it as $[w]_i  =\{w^i_1, ..., w^i_{n_i}\}$ where the words come from a $D$-sized vocabulary (usually $D = 8,000$). Neural CF can be extended to leverage text and then the interaction function has the form of $f(u,i,[w]_i | \Theta)$.

Based on neural architectures, it is flexible to enable us to exploit interaction data and content information seamlessly via centralized and local memories introduced in the following subsections (Section~\ref{paper:centralized} and ~\ref{paper:local}).

\subsection{Architecture}

\begin{figure}[!ht]
	\centering
    \includegraphics[height=4.0in,width=8.1cm]{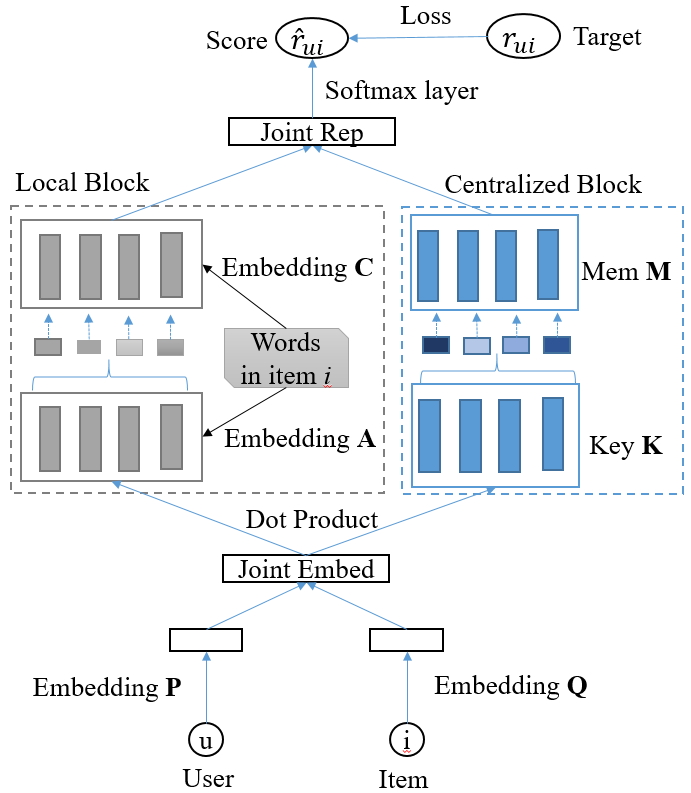}
	\caption{The Proposed Architecture of Local and Centralized Memories Recommender (LCMR) Model. In this illustration, there is one block for each of the local and centralized memories modules and the the size of the memories is four.}
	\label{fig:lcmr}
\end{figure}

Our contributions are summarized as follows:
\begin{itemize}
	\item Introducing an alternative approach to integrate item content via local memories;
	\item Proposing a novel recommender system LCMR which exploits interaction data and content information seamlessly in an end-to-end neural framework;
	\item Evaluating LCMR extensively on real-world datasets and conducting analyses to help understand the impact of local and centralized memories for LCMR.
\end{itemize}

The architecture for the proposed LCMR model is illustrated in Figure~\ref{fig:lcmr}. In general, besides the layers of input, embedding, and output, LCMR is stacked by multiple building blocks to learn (highly nonlinear) interaction relationships between users and items. The building blocks consist of local and centralized memories. The information flow in LCMR goes from the input $(u,i)$ to the output $\hat{r}_{ui}$ through the following layers:
\begin{itemize}
	\item {\bf Input layer: $(u,i) \rightarrow \bm{x}_u,\bm{x}_i$} This module encodes user-item interaction indices. We adopt the one-hot encoding. It takes user $u$ and item $i$, and maps them into one-hot encodings $\bm{x}_u \in \{0,1\}^{m}$ and $\bm{x}_i \in \{0,1\}^{n}$ where only the element corresponding to that index is 1 and all others are 0.
	\item {\bf Embedding lookup: $\bm{x}_u,\bm{x}_i \rightarrow \bm{x}_{ui}$} This module embeds one-hot encodings into continuous representations and then concatenates them as
	\begin{equation}\label{eq:concat}
		\bm{x}_{ui} = [\bm{P}^T\bm{x}_u, \bm{Q}^T\bm{x}_i]
	\end{equation}
	to be the input of following building blocks.
	\item {\bf Multi-hop centralized blocks: $\bm{x}_{ui} \rightsquigarrow \bm{z}_{ui}^c$} This module is a pure CF approach to exploit user-item interaction data. It takes the continuous representations from the embedding module and then transforms to a final latent representation:
    \begin{equation}
      \bm{z}_{ui}^c= \phi^c(\bm{x}_{ui}),
    \end{equation}
    where $\phi^c$ denotes the computing function in the module.
    \item {\bf Multi-hop local blocks: $\bm{x}_{ui} \rightsquigarrow \bm{z}_{ui}^l$} This module is an extended CF approach to integrate the item content with the guidance of interaction data. The item content is modelled by memories. It takes both representations from the embedding module and text associated with the item to a final latent representation:
    \begin{equation}
      \bm{z}_{ui}^l= \phi^l(\bm{x}_{ui}, [w]_i),
    \end{equation}
    where $\phi^l$ denotes the computing function in the module.
	\item {\bf Output layer: $[\bm{z}_{ui}^c, \bm{z}_{ui}^l] \rightarrow \hat r_{ui}$} This module predicts the score $\hat r_{ui}$ for the given user-item pair based on the concatenated representation:
\begin{equation}\label{eq:joint-rep}
 \bm{z}_{ui} = [\bm{z}_{ui}^c, \bm{z}_{ui}^l]
\end{equation}
from the multi-hop blocks. The output is the probability that the input pair is a positive interaction. This can be achieved by a softmax/logistic layer $\phi_o$:
	\begin{equation}
		\hat r_{ui} = \phi_o(\bm{z}_{ui}) = 1 / (1 + \exp(-\bm{h}^T \bm{z}_{ui})),
	\end{equation}
	where $\bm{h}$ is the parameter.
\end{itemize}

The centralized memory module is mainly to learn (highly) nonlinear representations from user-item interactions through multiple nonlinear transformations. The local memory module is mainly to learn text semantics from item content with the guidance of user-item interactions. A module consists of multi-hop blocks. In a unified model, LCMR achieves to learn both interaction representations and content semantics by jointly training centralized and local memory modules.

\subsection{Centralized Memory Module: Exploiting Interactions}\label{paper:centralized}

The basic idea of neural CF approaches is to learn the highly complex matching relationships between users and items from their interactions data by multiple non-linear transformations usually implemented by neural networks~\cite{dziugaite2015neural,he2017neural}. Inspired by the latent relational metric learning (LRML) model~\cite{Tay2018latent} which showed that user-item specific latent relations can be generated by a centralized memory module, we similarly design a centralized memory module\footnote{The input to our centralized memory module is different from that in the LRML model. And our module contains multiple memory blocks while LRML has only one.} to learn a latent representation from the joint user-item embedding: $\bm{z}_{ui}^c= \phi^c(\bm{x}_{ui})$.

A centralized (or global) memories block (illustrated by a dotted rectangle box marked as `Centralized Block' in Figure~\ref{fig:lcmr}) is parameterized by a memory matrix $\bm{M}^c \in \mathbb{R}^{N \times d}$ and a key matrix $\bm{K}^c \in \mathbb{R}^{N \times d}$, where $N$ is the size of the memories and $d$ is the dimensionality of embeddings, shared by all user-item interaction data. The first block takes the concatenated embeddings of users and items as its input, and then produces the transformed representations as its output which in turn is the input to the next block if there are multi-hops. For simplicity, we show the computing path in the module which has only one block in the following.

The computation path $\phi^c$ goes through $\bm{x}_{ui} \overset{\bm{K}^c} {\rightarrow}{\bm{a}} \overset{\bm{M}^c}{\rightarrow} {\bm{z}^c_{ui}}$. Firstly, we perform a content-based addressing mechanism between $\bm{x}_{ui}$ and $\bm{K}^c_j$ (the $j$-th row vector of $\bm{K}^c$) by a similarity measure $g(\cdot, \cdot)$ (a dot product) to produce unnormalized weights:
\begin{equation}\label{eq:attention-basic}
\bm{a}_j^{\text{un}} = g(\bm{x}_{ui}, \bm{K}^c_j), \quad j = 1, ..., N.
\end{equation}
The unnormalized attentive weights are normalized to be a simplex vector, e.g. by a softmax function:
\begin{equation}\label{eq:attention-normal}
\bm{a}_j = \frac{\exp({\beta \bm{a}_j}^{\text{un}})}{\sum_k \exp({\beta \bm{a}_k}^{\text{un}})},
\end{equation}
where the parameter $\beta$ has two functions. It can stabilize the numerical computation when the  exponentials of the softmax function are very large, e.g. the dimension $d$ is too high~\cite{vaswani2017attention}. We can also use it to amplify or attenuate the precision of the attention~\cite{graves2014neural}. We set $\beta = d^{-\frac{1}{2}}$ by scaling along with the dimensionality.

Secondly, we use the attentive weights to compute a weighted sum over memories $\bm{M}^c$ as the output:
\begin{equation}\label{eq:weighted-sum}
\bm{z}^c_{ui} = \sum\nolimits_j \bm{a}_j \bm{M}^c_j.
\end{equation}

Usually there are multiple nonlinear transformations in a neural architecture. Each block has its own memory matrices $\bm{K}^c$ and $\bm{M}^c$. The parameters in the centralized module are $\theta^c_f = \{ \{\bm{K}^c\}, \{\bm{M}^c\} \}$.

\subsection{Local Memory Module: Integrating Text}\label{paper:local}

Items are usually associated with the content information such as unstructured text (e.g., abstracts of articles and reviews of products). CF approaches can be extended to exploit the content information~\cite{wang2011collaborative,bansal2016ask} and user reviews~\cite{mcauley2013hidden,zheng2017joint}. Memory networks can reason with an external memory~\cite{weston2016memory,sukhbaatar2015end}. Due to the capability of naturally learning word embeddings to address the problems of word sparseness and semantic gap, we design a local memory module to naturally model item content via memories. These memories are dynamic (or local) since they are specific to the input query and attentively learn what to exploit with the guidance of user-item interaction.

A local (or dynamic) memories block is illustrated by a dotted rectangle box marked as `Local Block' in Figure~\ref{fig:lcmr} where the affiliated irregular rectangle box donotes the external memories containing words in item $i$. For simplicity, in the following, we only show one block in the computing path.

Suppose we are given words (coming from a $D$-sized vocabulary) $w_1,...,w_j$ in item $i$ as the input to be stored in the memory. The entire set of $\{w_j\}$ are converted into key vectors $\{\bm{K}^l_j\}$ of dimension $d$ computed by embedding each $w_j$ in a continuous space using an embedding matrix $\bm{A}$ (of size $d \times D$). The query is the concatenated user-item embeddings $\bm{x}_{ui}$. In the embedding space, we compute the match between query $\bm{x}_{ui}$ and each memory $\{\bm{K}^l_j\}$ by taking the inner product followed by a softmax function, similar to Eq.(\ref{eq:attention-basic}) and Eq.(\ref{eq:attention-normal}) respectively. Each $w_j$ has a corresponding memory vector $\{\bm{M}^l_j\}$ by another embedding matrix $\bm{C}$ (of size $d \times D$). The resulting representation $\bm{z}^l_{ui}$ is then a sum over memories $\{\bm{M}^l_j\}$ weighted by attentive probabilities, similar to Eq.(\ref{eq:weighted-sum}).

Usually there are multiple nonlinear transformations in a neural architecture. The embedding matrices $\bm{A}$ and $\bm{C}$ in each block are shared. The parameters in the local memories module are $\theta^l_f = \{ \bm{A}, \bm{C} \}$.

\subsection{Optimization and Learning}

Due to the nature of the implicit feedback and the task of item recommendation, the squared loss $(\hat r_{ui} - r_{ui})^2$ may be not suitable since it is usually for rating prediction. Instead, we adopt the binary cross-entropy loss:
\begin{equation}
\mathcal{L} =  - \sum\nolimits_{(u,i) \in \mathcal{S} } r_{ui} \log{\hat r_{ui}} + (1-r_{ui}) \log(1 - \hat r_{ui} ),
\end{equation}
where $\mathcal{S} = \bm{R}^+ \cup \bm{R}^-$ are the union of observed interaction matrix and randomly sampled negative pairs. This objective function has a probabilistic interpretation and is the negative logarithm likelihood of the following likelihood function:
\begin{equation}
L(\Theta | \mathcal{S} ) = \prod\nolimits_{(u,i) \in \bm{R}^+} \hat r_{ui} \prod\nolimits_{(u,i) \in \bm{R}^-} (1 - \hat r_{ui}),
\end{equation}
where the model parameters are:
$$\Theta=\{\bm{P}, \bm{Q}, \bm{h}, \theta^l_f, \theta^c_f\}.$$
This objective function can be optimized by the stochastic gradient descent (SGD) algorithm and its variants.

Note that the local and centralized modules in LCMR are jointly training and there is a distinction from ensemble learning. In an ensemble, predictions of individual models are combined only at the inference process but not at the training process. In contrast, joint training optimizes all parameters simultaneously by taking both the local and centralized modules into account at the training process. Furthermore, the entire model is more compact by sharing the user and item embeddings.

\section{Experiments}

In this section, we conduct empirical study to answer the following questions: 1) how does the proposed LCMR model perform compared with state-of-the-art recommender systems; and 2) how do local and centralized memories contribute to the proposed framework. We firstly introduce the evaluation protocols and experimental settings, and then we compare the performance of different recommender systems. We further analyze the LCMR model to understand the impact of the two memories modules, followed by showing the optimization curves.

\subsection{Experimental Settings}

\noindent
{\bf Datasets} We conduct experiments on two datasets. The first dataset,  CiteULike\footnote{\url{http://www.cs.cmu.edu/~chongw/data/citeulike/}}, is widely used to evaluate the performance on scientific article recommendation~\cite{wang2011collaborative}. The second dataset, Company Mobile, provided by a company %Cheetah Mobile\footnote{\url{http://www.cmcm.com/en-us/}}
is on the domain of news reading in the region of New York City, USA in one month (January 2017). The other information, such as dwell time, publisher, and demographic data, is not used in this paper. For CiteULike, we use the version released in the work~\cite{wang2011collaborative}, and the size of vocabulary is 8,000 and there are about 1.6M words. For Company Mobile dataset, we preprocess it following the work~\cite{wang2011collaborative}. We removed users with fewer than 10 feedback. For each item, we use only the news title. We filter stop words and use tf-idf to choose the top 8,000 distinct words as the vocabulary. This yields a corpus of 0.6M words. The statistics of datasets are summarized in Table~\ref{tb:data}. Both datasets are sparser than 99\%. Note that CiteULike is long text of paper abstracts (the number of average words per item is 93.5), while Company Mobile is short text of news titles (the number of average words per item is 6.7).

\begin{table}[]					
\centering
\caption{Datasets and Statistics. }
\label{tb:data}			
\resizebox{0.45\textwidth}{!}{					
    \begin{tabular}{c | cc}
\hline \hline
    Dataset             & CiteULike     & Company Mobile \\
\hline \hline
\#Users                 &   5,551       &    18,387    \\
%\hline
\#Items                 &   16,980      &    92,008     \\
%\hline
\#Feedback              &   204,986    &    569,749      \\
\#Words                 &   1,587,000   &     612,839       \\
%\hline
Rating Density (\%)     &   0.218        &    0.034    \\
%\hline
%\hline
Avg. Words per Item     &    93.5       &      6.7    \\
\hline  \hline
\end{tabular}
}										
\end{table}

\noindent
{\bf Evaluation protocols} For item recommendation task, the leave-one-out (LOO) evaluation is widely used and we follow the protocol in the work of neural collaborative filtering~\cite{he2017neural}. That is, we reserve one interaction (usually the latest one or randomly picked if no temporal information) as the test item for each user. We follow the common strategy which randomly samples 99 (negative) items that are not interacted by the user and then evaluate how well the recommender can rank the test item against these negative ones. The performance is measured by Hit Ratio and Normalized Discounted Cumulative Gain (NDCG), where the ranked list is cut off at 10. The former metric measures whether the test item is present on the top-10 list and the latter also accounts for the hit position by giving higher reward for top ranks. Results are averaged over all test users. The higher the values, the better the performance.

\noindent
{\bf Baselines} We compare with various baselines. The first class methods are non-personalized. {\bf ItemPOP} ranks items by their popularity, that is, the number of interacted users. The second class methods are pure CF. {\bf BPRMF}, Bayesian personalized ranking~\cite{rendle2009bpr}, optimizes factorization with a pairwise ranking loss function rather than the pointwise as we did, which is tailored to learn from the implicit feedback. It is a state-of-the-art traditional CF technique. {\bf MLP}, multilayer perceptron~\cite{he2017neural}, learns the nonlinear interaction function using feedforward neural networks. The last class methods are extended CF. We compare with {\bf CTR}, collaborative topic regression~\cite{wang2011collaborative}, combines MF and topic modeling. It is a state-of-the-art model which also exploits auxiliary text sources. Actually, the CiteULike dataset is introduced in the CTR paper.

\begin{table*}[]													
\centering													
\caption{Results of Hit Ratio and NDCG ($\times 100$) at the cut-off 10. The last column is the relative improvement of LCMR vs the best baseline (marked by a star (*)). The \textbf{best} scores are boldfaced.}	
\label{tb:result}													
%\resizebox{0.85\textwidth}{!}{
\begin{tabular}{cc c c c c c | c}		
\hline \hline											
                 Dataset     & Metric & ItemPOP & BPRMF  & MLP  & CTR & LCMR & Improvement\\
\hline 	\hline						
\multirow{2}{*}{CiteULike}   &Hit Ratio & 27.35 & 74.43 & 78.02 & 83.05* & {\bf 84.60} & 1.87\%\\
 \cline{2-8}
                              & NDCG  & 18.32  &49.44  &51.23  & 57.79* & {\bf 61.07} & 5.68\%  \\	
\hline \hline			
\multirow{2}{*}{Company Mobile} &Hit Ratio & 67.91 & 68.39 & 73.20* & 68.23 & {\bf 75.74} & 3.47\% \\	
\cline{2-8}
                               & NDCG & 45.55 & 50.74 & 51.80* &  46.34 & {\bf 54.62} & 5.44\%  \\	
\hline \hline	
\end{tabular}
%}													
\end{table*}

\noindent
{\bf Settings} For BPRMF, we use the implementation of LightFM\footnote{\url{https://github.com/lyst/lightfm}} which is a widely used CF library in various competitions. For neural CF methods, we use the implementation released by the authors\footnote{\url{https://github.com/hexiangnan/neural_collaborative_filtering}}. For CTR, we use the implementation released by the authors\footnote{\url{http://www.cs.cmu.edu/~chongw/citeulike/}}. Our method is implemented using TensorFlow\footnote{\url{https://www.tensorflow.org}} running on the Nvidia GPU GTX TITAN X. As a general setting, parameters are randomly initialized from Gaussian $\mathcal{N}(0, 0.01^2)$. The optimizer is adaptive moment estimation (Adam)~\cite{kingma2014adam} with initial learning rate 0.001. The size of mini batch is 128. The ratio of negative sampling is 1. We tune hyper-parameters $\Phi = \{L,N,d$\} (hops $L$, size of memories $N$, and dimensionality $d$) on the validation set. Best results are reported on the test set during 50 epochs where parameters $\Phi$ are fixed corresponding to the best validation performance (default values: $L=3, N=100, d=200$).

\subsection{Comparisons of Different Recommender Systems}												

The comparison results are shown in Table~\ref{tb:result} and we have the following observations.

Firstly, LCMR outperforms the traditional CF method BPRMF on the two datasets in terms of both Hit Ratio and NDCG. On CiteULike, LCMR obtains a large improvement in performance gain with relative 13.66\% Hit Ratio and 23.52\% NDCG. On Company Mobile, LCMR obtains a large improvement in performance gain with relative 10.75\% Hit Ratio and 7.65\% NDCG. Compared with the traditional matrix factorization based models where the dot product is used to match user and item, the results show the benefit of learning nonlinear interaction function through multiple nonlinear transformations.

Secondly, LCMR also outperforms the neural CF method MLP on the two datasets in terms of both Hit Ratio and NDCG. On CiteULike, LCMR obtains a large improvement in performance gain with relative 8.43\% Hit Ratio and 19.21\% NDCG. On Company Mobile, LCMR still obtains reasonably significant improvements with relative 3.47\% Hit Ratio and 5.44\% NDCG. Compared with pure neural CF methods which exploit the interaction data only, the results show the benefit of integrating text information through local memories module of LCMR.

Lastly, LCMR outperforms the extended CF method CTR by a large margin on Company Mobile dataset with relative 11.01\% Hit Ratio and 17.87\% NDCG; while LCMR still obtains reasonably significant improvements on CiteULike dataset with relative 1.87\% Hit Ratio and 5.68\% NDCG. Note that the Company Mobile dataset is short text of news titles and there is difficulty in learning topic distributions for the CTR model. Furthermore, the news articles have the timeliness and hotness characteristics and it may explain that the non-personalized method ItemPOP is a competitive baseline.

In summary, the empirical results of LCMR demonstrate the superiority of local and centralized memory modules to exploit the interaction and text information. In the following subsection, we investigate the contributions from the components of LCMR.

\subsection{The Impact of Local and Centralized Memories Modules}

\begin{figure}
    \centering
    \includegraphics[width=6.5cm, height=4.9cm]{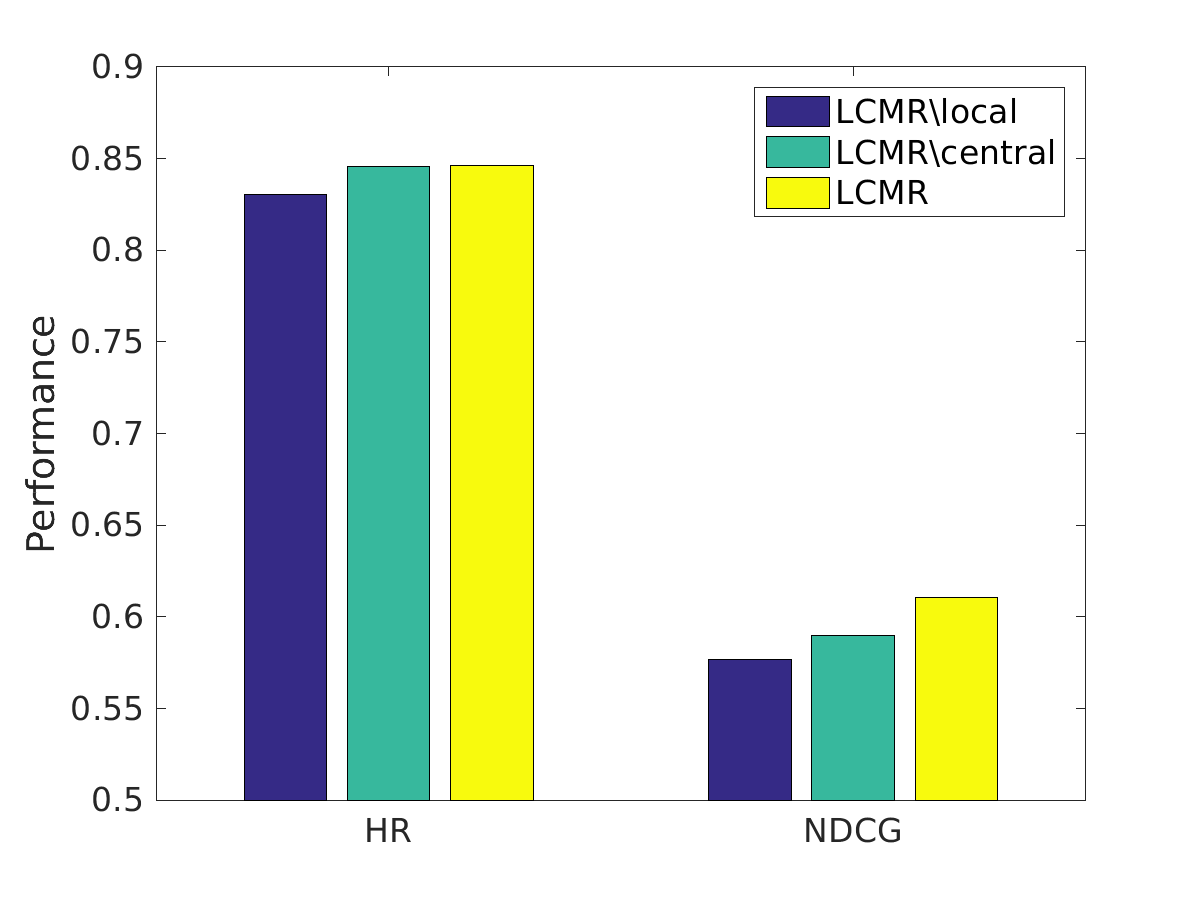}
    \caption{Impact of Local and Centralized Memories Modules of the LCMR model on CiteULike}
    \label{fig:memories}
\end{figure}

We have shown the effectiveness of local and centralized memories modules in our proposed LCMR framework. We now investigate the contribution of each memory module to the LCMR by eliminating the impact of local and centralized modules from it in turn\footnote{If we eliminate both the centralized and local memory modules, then LCMR computes $\hat r_{ui} = 1 / (1 + \exp(-\bm{h}^T \bm{x}_{ui}))$ where $\bm{x}_{ui} = [\bm{P}^T\bm{x}_u, \bm{Q}^T\bm{x}_i]$ and model parameters $\Theta=\{\bm{h}, \bm{P},\bm{Q} \}$, which is similar to the matrix factorization methods like BPRMF but has an extra nonlinear sigmoid transformation.}:
\begin{itemize}
\item {\bf LCMR$\backslash$local}: Eliminating the impact of local memory module by setting $\bm{z}_{ui} = \bm{z}_{ui}^c$ in Eq.(\ref{eq:joint-rep}); that is, removing the local multi-hop blocks. The model parameters $\Theta=\{\bm{P}, \bm{Q}, \bm{h}, \theta^c_f\}.$
\item {\bf LCMR$\backslash$central}: Eliminating the impact of centralized memory module by setting $\bm{z}_{ui} = \bm{z}_{ui}^l$ in Eq.(\ref{eq:joint-rep}); that is, removing the centralized multi-hop blocks. The model parameters $\Theta=\{\bm{P}, \bm{Q}, \bm{h}, \theta^l_f\}.$
\end{itemize}

The comparison results of LCMR and its two modules on CiteULike are shown in Figure~\ref{fig:memories}. The performance degrades when either local or centralized memories modules are eliminated. In detail, LCMR$\backslash$local and LCMR$\backslash$central reduce 5.85\% and 3.57\% relative NDCG performance respectively, suggesting that both local and centralized memories contain essential information for recommender. Naturally, removing the local memory module degrades performance worse than removing the centralized memory module due to the losing of text information source.

\subsection{Sensitivity to Embedding Dimensionality}

\begin{figure}
    \centering
    \includegraphics[width=6.5cm, height=4.9cm]{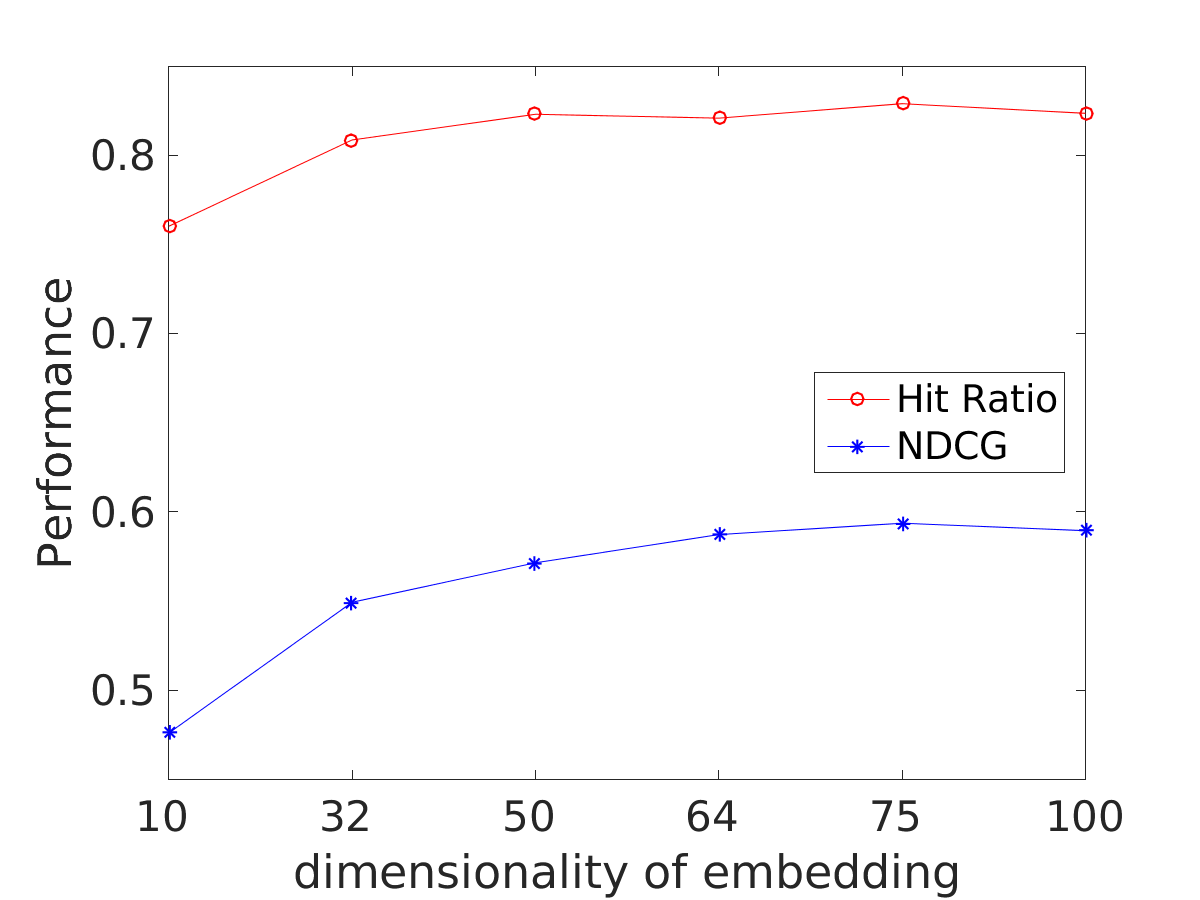}
    \caption{Performance with Dimensionality of Embedding on CiteULike}
    \label{fig:embedding}
\end{figure}

The dimensionality of the joint embeddings, i.e. $\bm{x}_{ui} = [\bm{P}^T\bm{x}_u, \bm{Q}^T\bm{x}_i]$ in Eq.(\ref{eq:concat}), controls the model complexity. Figure~\ref{fig:embedding} shows the sensitivity of our model to it on CiteULike. Note that The $x$-axis in Figure~\ref{fig:embedding} equals half of the dimensionality of the joint embeddings. In other words, it is the dimensionality of user (item) embeddings. It clearly indicates that the embedding should not be too small due to the possibility of information loss and the limits of expressiveness. It can get good results when the joint dimensionality is around 150.

\subsection{Optimization and Running Time}

\begin{figure}
    \centering
    \includegraphics[width=6.8cm, height=5.8cm]{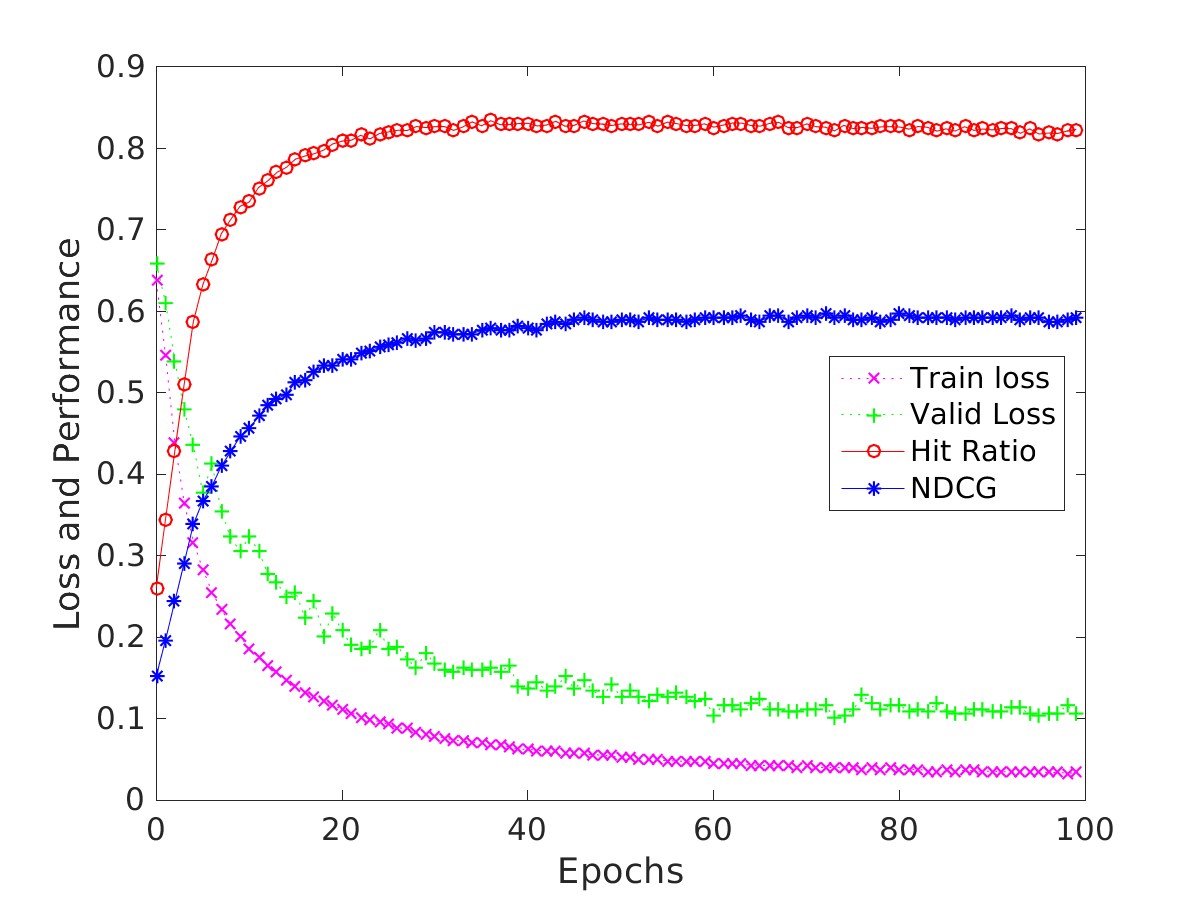}
    \caption{Optimization Curves of Performance and Loss with Iterations on CiteULike}
    \label{fig:epoch}
\end{figure}

We show optimization curves of performance and loss (averaged over all examples) against iterations on CiteULike in Figure~\ref{fig:epoch}. The model learns quickly in the first 20 iterations, improves slowly until 30, and stabilizes around 50, though losses continue to go down. The average time per epoch takes 64.9s and as a reference, it is 34.5s for MLP.

\section{Related Works}

Recently, neural networks have been proposed to parameterize the interaction function between users and items. The MF Autoencoder~\cite{vanBaalen2016autoencoding} and NNMF model~\cite{dziugaite2015neural} parameterize the interaction function by a multilayer FFNN. The MLP~\cite{he2017neural} and Deep MF~\cite{xue2017deep} also use FFNNs. The basic MLP architecture is extended to regularize the factors of users and items by social and geographical information~\cite{yang2017bridging}. Other neural approaches learn from the explicit feedback for rating prediction task~\cite{sedhain2015autorec,zheng2017joint,catherine2017transnets,wu2017recurrent}. We learn from the implicit feedback for top-N recommendation~\cite{cremonesi2010performance,wu2016collaborative}.

Additional sources of information are integrated into CF to alleviate the data sparsity issues. Neural networks have been used to extract the features from auxiliary sources such as audio~\cite{van2013deep}, text~\cite{wang2015collaborative,kim2016convolutional,huang2016transferring,bansal2016ask}, image~\cite{he2016vbpr,chen2017attentive}, and knowledge base~\cite{zhang2016collaborative}. As for the interaction data, these works rely on matrix factorization to model the user-item interactions. We learn the interaction function and exploit auxiliary sources jointly under a generic neural architecture by the modules of centralized and local memories.

\section{Conclusion}

We proposed a novel neural architecture, LCMR, to jointly model user-item interactions and integrate unstructured text for collaborative filtering with implicit feedback. By modeling text content as local memories, LCMR can attentively learn what to exploit from the unstructured text with the guidance of user-item interaction. LCMR is a unified framework as it embraces pure CF approaches and CF with auxiliary information. It shows better performance than traditional, neural, and extended approaches on two datasets under the Hit Ratio and NDCG metrics. Furthermore, we conducted ablation analyses to understand the contributions from the two memory components. We also showed the optimization curves of performance and loss.

The datasets contain other information such as tags of items and profiles of users which can be exploited to alleviate the cold-start issues in the future works. Besides basing on the memory networks, other kinds of neural networks like recurrent and convolutional networks may be used.

\bibliographystyle{named}
\bibliography{ijcai18}

\end{document}